\documentclass[10pt]{iopart}

\usepackage{graphicx}% Include figure files
\usepackage{bm}% bold math

\begin{document}

\title{Laser frequency stabilization to a single ion}
\author{Ekkehard Peik, Tobias Schneider and Christian Tamm}
\address{Physikalisch-Technische Bundesanstalt,
Bundesallee 100, 38116 Braunschweig, Germany}
\ead{ekkehard.peik@ptb.de}

\begin{abstract}
A fundamental limit to the stability of a single-ion optical frequency standard is set by quantum noise in the  measurement of the internal state of the ion.  
We discuss how the interrogation sequence and the processing of the atomic resonance signal can be optimized in order to obtain the highest possible stability under realistic experimental conditions. A servo algorithm is presented that  stabilizes a laser frequency to the single-ion signal and that eliminates errors due to laser frequency drift. Numerical simulations of the servo characteristics are compared to experimental data from a frequency comparison of two single-ion standards based on a transition at 688 THz in $^{171}$Yb$^+$. Experimentally, an instability $\sigma_y(100\,{\rm s})=9\cdot 10^{-16}$ is obtained in the frequency difference between both standards.
\end{abstract}

\submitto{J. Phys. B: Atomic, Molecular \& Optical Physics, 13 July 2005}
\maketitle

\section{Introduction}

Optical atomic frequency standards based on single trapped and laser-cooled ions have made important progress and
may soon reach and surpass the level of accuracy that is achievable with cesium atomic clocks in the microwave region today \cite{6symp,gill,nistrev,bize,peik,margolis,oskay}.
An ion trap provides a very well-controllable environment and allows one to work at the conceptually simplest level: with one single atom. It is this simplicity that provides the basis for the high accuracy that can be obtained in spectroscopic measurements on these systems.  

In atomic frequency standards the frequency of an oscillator producing an electromagnetic wave is stabilized to 
an atomic reference transition. The error signal for the frequency lock 
is derived by modulating the oscillator frequency around the atomic resonance and by measuring the resulting modulation of the frequency-dependent excitation probability $p$ to the upper atomic level.
If only a single ion is used, 
a difficulty arises from the weakness of this signal and the consequently limited signal-to-noise ratio.
The ion is probed repeatedly by pulses of the radiation.
The state of the ion after an excitation attempt can be determined by applying laser radiation to induce resonance fluorescence on a transition that shares the lower state with the reference transition. 
This scheme was proposed by Dehmelt and is sometimes called electron shelving \cite{deh}.
Let us denote the two levels that are connected by the reference transition
as $|A\rangle$ and $|B\rangle$ and assume that the ion is initially in the lower state $|A\rangle$.
After an excitation attempt the ion generally will be in a superposition state
$\alpha |A\rangle + \beta |B\rangle$ and the measurement with the electron shelving scheme is equivalent to determining the eigenvalue $P$ of the projection operator $\hat P=|B\rangle\langle B|$. 
If no fluorescence 
is observed (the probability for this outcome being $p=|\beta|^2$) the previous excitation attempt is regarded  successful ($P=1$), whereas the observation of fluorescence indicates that the excited state was not populated ($P=0$).
In one measurement cycle only one binary unit of spectroscopic information is obtained. Under conditions where the average excitation probability $p$ is $0.5$, the result of a sequence of cycles is a random sequence of zeros and ones and the uncertainty in a prediction on the outcome of the next cycle is always maximal.
These population fluctuations and their relevance in atomic frequency standards were first discussed by Itano et al., who named the phenomenon quantum projection noise (QPN) \cite{qpn}.
A simple calculation shows that the variance of the projection operator is given by \cite{qpn}
\begin{equation}
(\Delta \hat P)^2=p(1-p).	
\end{equation}
If the signal-to-noise ratio of a state measurement with a single atom is estimated as ${\rm SNR}=p/\sqrt{p(1-p)}$, it can be seen that for $p=0.5$, ${\rm SNR}=1$ only.

The stability and the accuracy of an atomic frequency standard are ultimately determined by the signal-to-noise ratio,
by the linewidth $\Delta \nu$ and by the fluctuations of the center frequency $\nu_0$ of the atomic resonance \cite{vanier}.
The instability that is obtained after an averaging time $T$ is conveniently expressed by the Allan deviation $\sigma_y(T)$ which  can be written as
\begin{equation}
\sigma_y(T) =\frac {C}{{\rm SNR}} \frac{\Delta \nu}{\nu_0}\sqrt{\frac{t_c}{T}}. 
\end{equation}
This expression describes the case of a standard that is operated in interrogation cycles of duration $t_c$.
${\rm SNR}$ denotes the signal-to-noise ratio that is obtained in one cycle and $C$ is a numerical constant of order unity.
The instability decreases inversely 
proportional to the square root of the number of cycles,
giving rise to $\sigma_y(T)\propto 1/\sqrt{T}$.
Parameters that are obtainable with present single-ion optical frequency standards \cite{bize,peik,margolis} are 
of the order of magnitude
$\Delta \nu / \nu_0=10^{-15}$ and $t_c=1$~s. There are prospects to reach an accuracy in the range of $10^{-18}$ \cite{deh,nistrev}, but from equation (2) it can be seen that with $C/{\rm SNR}\approx 1$, an averaging time of about $10^6$~s or roughly 10 days would be required to complete a single frequency measurement at this accuracy. 
For the practical use of such a standard it will therefore be of high importance to find the interrogation conditions
and servo techniques which minimize $C/{\rm SNR}$, i.e. which make the most efficient use of the single-ion signal in order to provide a stable reference frequency.  

In this paper we present a theoretical and experimental study of the stability of a single-ion frequency standard under conditions where QPN is the dominant noise source.
Because QPN presents a fundamental noise limit if only a single ion is probed, it defines the target below which technical noise due to the short-term frequency instability of the probe laser should be reduced.
In the first part of the paper we present results of numerical calculations that allow one to find optimized excitation parameters for the case that the
linewidth of the laser is smaller than the natural linewidth of the atomic reference transition.   
The choice of this regime is motivated by the fact that most optical frequency standards presently use reference transitions with natural linewidths in the range of a few hertz to a few hundred hertz \cite{6symp}, whereas the most advanced laser sources already have subhertz linewidths \cite{young}. With the practically unlimited interaction time 
for the trapped ion, this situation is in contrast to atomic microwave frequency standards where 
the natural linewidth is usually much smaller than the linewidth of the interrogation oscillator and where the resolution 
is  limited by the finite interaction time. We compare the instability that is obtainable by excitation with a single laser pulse
(Rabi excitation) and with two pulses (Ramsey excitation).  The influence of a non-zero laser linewidth and  of cycle dead time that is needed for the preparation of the ion and for state detection is discussed. 
While a final optimization will also have to take the specific frequency noise properties of the probe laser into account, these general results are meant to provide initial guidance in the selection of suitable experimental parameters.
During the preparation of the manuscript two related studies were published  
\cite{champ,riis} and we will comment on the relation to that work at the end of section (3).

In the second part of the paper we present a servo algorithm that can be used to stabilize the laser frequency to the single-ion signal. We investigate the dynamic behaviour of the servo system and the resulting stability of the frequency standard by numerical Monte Carlo simulations. Possible servo errors due to laser frequency drift are also considered.  
Results from the simulations are compared to data from experiments with two Yb$^+$ frequency standards
that implement this servo algorithm. 
Previously, brief descriptions of frequency stabilizations to a single ion were given in \cite{ionlock1,npl1}
and a more detailed discussion including analytical expressions for the expected instability was presented by Bernard et al. \cite{bern}.     

\section{Comparison of Rabi and Ramsey excitation}

Using numerical calculations, we study possible excitation conditions for the interrogation of the single-ion resonance in order to find parameters that yield a minimum instability. We make the assumption that the obtainable resonance linewidth and excitation probability are limited by the finite lifetime $\tau$ of the excited atomic state. This regime will be typical for the operation of many single-ion optical frequency standards that are presently studied \cite{peik,margolis,rafac,becker}.
In optimizing the excitation conditions, a compromise has to be found between the aims of obtaining a narrow linewidth and a high excitation probability. Because of their different dependence of the signal on the probe time $t$, we compare the two methods of Rabi excitation and Ramsey excitation.

In order to calculate the shape of the excitation spectrum we use the vector representation of the optical Bloch equations in the form:
\begin{eqnarray}
\dot{u} &=& \Delta\omega v - \frac{1}{2} (\Gamma+\Gamma_L)u\\
\dot{v} &=& -\Delta\omega u-\omega_R w- 	\frac{1}{2} (\Gamma+\Gamma_L)v\\
\dot{w} &=& \omega_R v- \Gamma (w-\frac{1}{2})
\end{eqnarray}
where $2w=2p-1$ is the population difference between the excited and the ground state and $u$ and $v$ are the real and the imaginary parts of the coherence. $\omega_R$ denotes the Rabi frequency. The detuning of the laser frequency $f$ from the atomic resonance $\nu_0$ determines $\Delta\omega=2\pi(f-\nu_0)$. The damping rate $\Gamma=1/\tau$ describes the relaxation through spontaneous atomic decay and the additional coherence relaxation rate $\Gamma_L$ is used to model the influence of the laser linewidth. $\Gamma_L$ describes a loss of coherence due to white laser frequency noise which would result in a Lorentzian excitation spectrum of full width at half maximum (FWHM) $\Gamma_L/2\pi$ for Rabi excitation under the conditions
$\omega_R \ll \Gamma_L,~\Gamma \ll \Gamma_L$, and $1/t \ll \Gamma_L$.

By assigning a fixed value of the Rabi frequency $\omega_R$ to the optical transition we neglect the influence of the finite temperature of the ion. This is justified as long as the product of the ion's mean vibrational quantum number $\langle n\rangle$ and the squared Lamb-Dicke parameter 
is small:
$\langle n \rangle k^2 \hbar / 2 m \Omega \ll 1$ ($k$: laser wavenumber, $m$: mass, $\Omega$: vibration frequency of the ion). This condition can be fulfilled by the use of a trap with high vibrational frequency or by cooling close to the vibrational ground state $\langle n \rangle \approx 0$. Otherwise, the $n$-dependence of the Rabi frequency will wash out population oscillations and will make it impossible to deterministically excite the ion to state $|B\rangle$  by the application of a resonant $\pi$-pulse 
\cite{champ}.

\subsection{Rabi excitation with a pulse of duration $t$}

For various values of the Rabi frequency $\omega_R$ and the pulse duration $t$, the Bloch equations were integrated numerically to calculate the maximum excitation probability $p_{\rm max}$ at resonance. The width (FWHM) of the line profile $\Delta \omega_0$ was determined by searching for the detunings where the excitation probability drops to $p_{\rm max}/2$.
We assume that these detunings are selected for the interrogation of the atomic line in the operation
of the frequency standard. 
The suitability of the excitation conditions can then be determined by calculating a dimensionless stability parameter that we define as
\begin{equation}
S_P=\sqrt{\frac{1}{2}p_{\rm max}(1-\frac{1}{2}p_{\rm max})}\frac{\Delta\omega_0 \tau}{p_{\rm max}}\sqrt{\frac{t+t_d}{\tau}}.	
\end{equation}
With this definition, the Allan deviation (cf. equation (2)) is approximately given by: 
\begin{equation}
\sigma_y(T)= S_P \frac{1}{2\pi \nu_0 \tau} \sqrt{\frac{\tau}{T}}.
\end{equation}
The first factor in $S_P$ describes QPN for the interrogation at the half-linewidth detuning (cf. equation (1)) and the second factor is inversely proportional to the slope of the obtainable error signal. The last factor describes the expenditure of time for one excitation cycle, $t_c=t+t_d$,
consisting of the probe time $t$ and the dead time $t_d$ that is needed for detecting the state of the ion, for laser cooling and the preparation of the next interrogation cycle. 
The atomic lifetime $\tau$ serves as the scaling parameter for time and frequency.
The numerical search for the minimum of $S_P$ was performed on a grid of values for $t$ and $\omega_R$ that was chosen to determine the optimized parameters with a relative uncertainty of 2\%.

\subsection{Ramsey excitation with two short pulses separated by a time $t$}

In the optimization of Ramsey excitation with two short pulses (of a duration that is much smaller than $t$), the Rabi frequency does not have to be treated as a free parameter because, irrespective of 
the relaxation rates $\Gamma$ and $\Gamma_L$, the choice of two $\pi/2$-pulses always ensures the highest signal contrast and lowest instability.
In analogy to equation (6), the stability parameter for the case of Ramsey excitation is calculated as
\begin{equation}
S_R=\sqrt{\bar{p}(1-\bar{p})}\frac{\Delta\omega_0 \tau}{p_{\rm max}-p_{\rm min}}\sqrt{\frac{t+t_d}{\tau}}.	
\end{equation}
Here, $p_{\rm max}$ and $p_{\rm min}$ are the maximum and minimum excitation probabilities near the central Ramsey fringe and $\bar{p}=(p_{\rm max}+p_{\rm min})/2$ is the average excitation probability.  The full fringe width 
$\Delta\omega_0$ is determined numerically as the frequency difference between resonance and the first minimum.

By replacing $S_P$ by $S_R$ in equation (7), the Allan deviation obtainable with Ramsey excitation can be calculated. 
Since the actual slope of the spectral profiles at the FWHM points enters $S_P$ and $S_R$ only in the approximation  
$\Delta p / \Delta \omega_0$, we can not expect to obtain exactly identical Allan deviations in experimental realizations of the Ramsey and Rabi schemes with $S_P=S_R$, but we expect the two systems to show comparable instabilities to within a few percent under this condition. 

\section{Optimized parameters for Rabi and Ramsey excitation} 

\begin{figure}[t]
\begin{center}
\includegraphics[height=10cm]{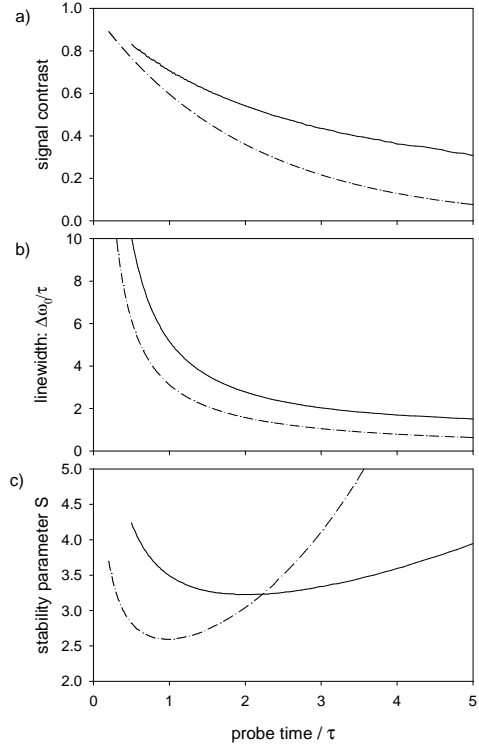}
\end{center}
\caption{Comparison of the obtainable stability with Rabi (solid line) and Ramsey (dot-dashed line) excitation for the case of negligible laser linewidth. The optimized signal contrasts $p_{\rm max}-p_{\rm min}$ (where $p_{\rm min}=0$ for Rabi excitation) (a), linewidths (b) and stability parameters (c) are plotted as a function of the probe time, measured in units of the atomic lifetime $\tau$.}
\end{figure} 

Using the stability parameters defined in equations (6,8) as figures of merit,
figure (1) shows the comparison of the optimized signal contrasts, linewidths, and stability parameters for
Rabi and Ramsey excitation.
It is assumed that $\Gamma_L$ is negligible compared to the natural decay rate $\Gamma$ and that the cycle dead time $t_d$
is much smaller than the probe time $t$. If the duration of the probe time is increased, a narrower resonance line is obtained but with a weaker signal. In the case of Ramsey excitation with two $\pi/2$-pulses, the signal contrast (figure (1a)) drops exponentially like 
\begin{equation}
p_{\rm max}-p_{\rm min}=e^{-t/2\tau}	
\end{equation}
while the resonance linewidth (figure (1b)) narrows continuously like $1/t$ and may reach values well below the natural linewidth $\Delta\nu_0=1/(2\pi\tau)$.
For Rabi excitation with an optimized pulse area close to the $\pi$-pulse condition $\omega_R t=\pi$,
the excitation probability decreases slower than exponentially (in the limit of weak continuous excitation one would expect $p_{\rm max} \propto \omega_R^2 $) but the linewidth never becomes smaller than $\Delta \nu_0$.
The optimization of the stability for each value of $t$ leads to pulse areas that are slightly larger than for a $\pi$-pulse. In the range $0.5<t/\tau<5$ the optimal pulse area can be approximately described by
$\omega_R t/\pi=1.071 + 0.068 \,t/\tau $. 

For Rabi excitation the lowest value of the stability parameter is $S_P=3.22$, obtained for $t=1.88\tau$
and $\omega_R=2.02 \Gamma$.
For Ramsey excitation $S_R=2.59$ can be achieved at $t=1.00\tau$, indicating an advantage of about 20\%
for the optimized stability  in comparison to the Rabi method.
The excitation spectra for these two cases are plotted in figure (2), showing that the optimization leads to
spectra with quite similar shapes of the central resonance.  
Using equation (7) these results indicate that for the parameters  of the Yb$^+$ frequency standard based on the $^2S_{1/2}\rightarrow {^2D}_{3/2}$ transition \cite{peik} with $\nu_0= 6.88 \cdot 10^{14}$~Hz and
$\tau=0.051$~s, an Allan deviation $\sigma_y(T)=2.7\cdot 10^{-15} /\sqrt{T/{\rm s}}$ should be achievable with Ramsey excitation.    

\begin{figure}[t]
\begin{center}
\includegraphics[height=5cm]{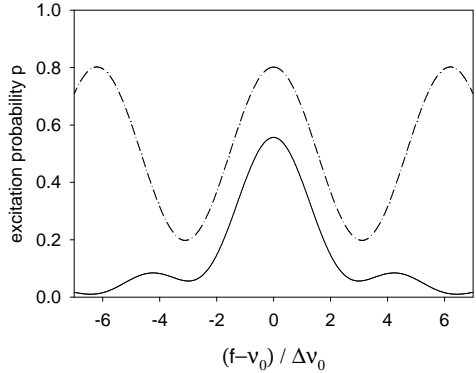}
\end{center}
\caption{Excitation spectra that yield optimal stability in the case of Rabi (solid line) and Ramsey excitation (dot-dashed line), corresponding to the minima of $S$ in figure (1c).}
\end{figure}

If realistic experimental conditions like non-zero laser linewidth and cycle dead time are taken into account,
the stability difference between the two excitation methods becomes less pronounced.
For the Ramsey scheme the additional relaxation $\Gamma_L$ due to the laser linewidth does not broaden the resonance but reduces the  fringe contrast.  In the case of Rabi excitation the line is broadened and the maximum excitation probability is reduced. In both cases it is advantageous to reduce the probe time $t$ below the values found for $\Gamma_L\ll 1/\tau$.
For a laser linewidth that is equal to the natural linewidth ($\Gamma_L/2\pi=\Delta\nu_0$) the optimized parameters shift to 
$t=1.10\tau$, $\omega_R=3.27\Gamma$, and $S_P=4.28$ for Rabi excitation, and to $t=0.50\tau$ and $S_R=3.64$ for Ramsey excitation, corresponding to degradations of the stability by 33\% and 40\%, respectively.
If the laser linewidth is larger than the natural linewidth, the probe time can be shortened further to adjust the Fourier transform limit to the laser linewidth.
In the limit $\Gamma_L/2\pi\gg \Delta\nu_0$ and when $\tau$ is replaced by $1/\Gamma_L$ in the definition of $S$, the optimization yields $t=2.44/\Gamma_L,~p_{\rm max}=0.78$, and $S_P=2.71$ for Rabi excitation, and
$t=1.00/\Gamma_L,~S_R=2.59$ for Ramsey excitation. The regime of a negligible natural linewidth is relevant if a highly forbidden reference transition like the $^2S_{1/2}\rightarrow {^2F}_{7/2}$ octupole transition in Yb$^+$ with its nanohertz natural linewidth is excited \cite{blythe}. 

The optimization problem that was discussed in this section has also been treated in two recent publications \cite{champ,riis} with emphasis on slightly different parameter regimes. Champenois et al. \cite{champ} also consider the influence of a non-zero vibrational quantum number of the ion and of high laser linewidth in the range $\Gamma_L/2\pi>\Delta \nu_0$. 
For the case considered here ($\langle n \rangle \approx 0$ and $\Gamma_L/2\pi\ll\Delta \nu_0$), they propose excitation profiles (figure (5) in \cite{champ}) that differ from the ones found here (figure (2)) and yield instabilities that are  about 50\% larger. In the work of Riis and Sinclair \cite{riis} the probe laser detuning that is used for the interrogation of the ion is always adjusted to the point where the slope of the excitation spectrum is maximal, and is not fixed to the half-linewidth detuning as in our calculation.
For Ramsey excitation, these two criteria agree and identical optimum excitation parameters are found.
The Allan deviation that is predicted in \cite{riis} on the basis of these parameters is about 35\% lower than that
calculated here using equations (7,8). This is due to the fact that our linear approximation 
$\Delta p/\Delta \omega_0$ for the slope of the Ramsey excitation spectrum in the definition of $S$ underestimates the maximum slope by a factor $2/\pi$.
For Rabi excitation and $\Gamma_L\ll 1/\tau$, the additional degree of freedom in reference \cite{riis} leads to an optimum with slightly shorter probe time ($t=1.63\tau$) and higher Rabi frequency ($\omega_R=2.19\Gamma$) than found by us.
The optimized Allan deviation in \cite{riis} is about 30\% lower than the value found here with equations (6,7).

\section{Influence of cycle dead time and the Dick effect}

\begin{figure}[t]
\begin{center}
\includegraphics[width=6cm]{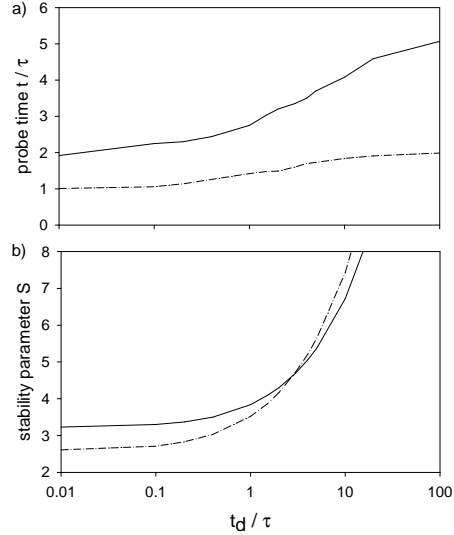}
\end{center}
\caption{Influence of cycle dead time $t_d$ on the stability for the cases of Rabi (solid lines) and Ramsey excitation (dot-dashed lines). With increasing dead time the optimized parameters shift to longer probe times (a). The minimum stability parameters (b) indicate that Rabi excitation becomes more favorable than Ramsey excitation for $t_d>3\tau$.}
\end{figure}

A non-negligible cycle dead time reduces the rate at which the spectroscopic information can be gathered. Considering this parameter in the optimization, it turns out to be advantageous to use longer probe times in order to prevent the ratio $t/t_c$ from becoming too small. 
Figure (3) shows how the optimized probe time increases and how the stability degrades with increasing dead time. For the case of long dead time ($t_d\gg \tau$), the optimized probe times converge to limits of $5.07\tau$ for Rabi excitation and $1.99\tau$ for Ramsey excitation, while the stability parameters diverge. Since for long probe times Rabi excitation results in a higher excitation probability (cf. figure (1a)), it provides a lower instability than the Ramsey method for dead times that are longer than $3\tau$.
From figure (3b) it can be seen that for an optimized
interrogation sequence the dead time should be kept below $\approx 0.2\,\tau$. 
In order to satisfy this requirement, the ion must be actively returned to the ground state 
after the state detection.

A further potentially harmful influence of the dead time on the 
stability comes from  the so-called Dick effect \cite{dickorg,dick}: the long-term stability of the standard may be degraded by down-conversion of the frequency noise of the interrogation oscillator at Fourier frequencies near the harmonics of $1/t_c$.
To evaluate the influence of this effect the noise spectrum of the oscillator has to be known.
An analytic estimate of the limiting
instability was given for a noise spectrum dominated by flicker frequency noise. For the case of Ramsey excitation one finds \cite{dick}:
\begin{equation}
\sigma_{y\,{\rm lim}}(T)\approx 	\frac{\sigma_{y\,{\rm osc}}}{\sqrt{2 \ln2}}\left| \frac{\sin(\pi t/t_c)}{\pi t/t_c}\right|\sqrt{\frac{t_c}{T}}
\end{equation}
where $\sigma_{y\,{\rm osc}}$ is the flicker floor instability of the oscillator.
With already achieved experimental parameters like $t/t_c>0.6$ and a flicker floor minimum 
$\sigma_{y\,{\rm osc}}<5\cdot 10^{-16}$ \cite{young}, it can be seen that the limitation from the Dick effect 
$\sigma_{y\,{\rm lim}}\approx 2\cdot 10^{-16}\sqrt{t_c/T}$
is well below the QPN-limited instability of $\sigma_y\approx 10^{-15}\sqrt{t_c/T}$ for present single-ion frequency standards.
   
\section{Servo algorithm}

A single interrogation of the ion provides only one binary unit of information
that has to be processed by a servo system before it can be used to act back on the laser frequency.
In this section we describe a suitable servo algorithm and investigate its dynamic properties.

Suppose the laser oscillates at a frequency $f$, close to the center of the resonance line.
A sequence of $2z$ cycles is performed in which the ion is interrogated 
alternately at the frequency $f_+=f+\delta_m$ and at $f_-=f -\delta_m$.  The 
numbers of successful excitations $n_+$ at $f_+$ and $n_-$ at $f_-$ are counted.
After an averaging interval of $2z$ cycles an error signal is calculated as
\begin{equation}
e= \delta_m \frac{n_+-n_-}{z},	
\end{equation}
and a frequency correction $g\cdot e$ is applied to the laser frequency before the next averaging interval is started:
\begin{equation}
f \rightarrow f+g\cdot e.	
\end{equation}
The numerical factor $g$ determines the dynamical response of the servo system  and we will call it `gain' for short.
Since the frequency corrections calculated for subsequent intervals are added up, this scheme realizes an integrating servo loop.

In order to obtain the maximum slope of the excitation spectrum, the best value for $\delta_m$ will be close to the half linewidth of the atomic resonance.
The time constant and the stability of the servo system are determined by the choice of the parameters $g$ and $z$.
If the laser frequency $f$ is initially one half linewidth below the atomic resonance and if $p_{\rm max}=1$, 
the resulting
value of $(n_+-n_-)/z$ will also be close to one. 
Consequently, with $g\approx 1$, the laser frequency will be corrected in a single step.
If $g\ll 1$, approximately $1/g$ averaging intervals will be required to bring the 
frequency close to the atomic resonance. For $g\approx 2$, one expects the laser frequency correction to jump between $-\delta_m$ and $+\delta_m$ without ever settling at a small value.  If the error signal is averaged over a large number of cycles $2z$, the response time of the servo
system is increased and the demands on the short-term stability of the probe laser become more stringent. 
For a small value of $z$,
however, the short-term stability of the system may be unneccessarily degraded by strong fluctuations in the error signal because of QPN. 

In order to guide the choice of suitable servo parameters, we performed  numerical Monte Carlo simulations of the servo action under the influence of QPN. As an approximation of the conditions obtained with Rabi excitation, an excitation spectrum of the form
\begin{equation}
p(f)=p_{\rm max} \,\left(\frac {\sin (\pi (f-\nu_0)/\gamma_0)}{\pi (f-\nu_0)/\gamma_0}\right)^2	
\end{equation}
was used. Here, $\gamma_0$ determines the linewidth ($\gamma_0\approx 1.13 \Delta \omega_0 /2\pi$) and $\delta_m=\gamma_0/2$ was chosen.
The time constant of the servo was determined from the exponential response of the error signal 
$e(t)\propto \exp(-t/t_{\rm servo})$
to a step-like frequency perturbation
and was found to be well described for $g/z<0.2$ by
\begin{equation}
t_{\rm servo}\approx 1.37  \frac{z}{g\cdot p_{\rm max}} t_c.	
\end{equation}
Since the time required to determine the error signal is $2zt_c$, the servo time constant can not be significantly shorter. With equation (14) this corresponds to a limit $g\cdot p_{\rm max}<1.46$. 
In fact, stable operation with a time constant of about $2t_c$ can be obtained for $g=1,~p_{\rm max}=1$ and $z=1$.
Damped oscillatory behaviour of the servo is observed to set in at $g\cdot p_{\rm max}\approx 1.5$. 

\begin{figure}[t]
\begin{center}
\includegraphics[width=7cm]{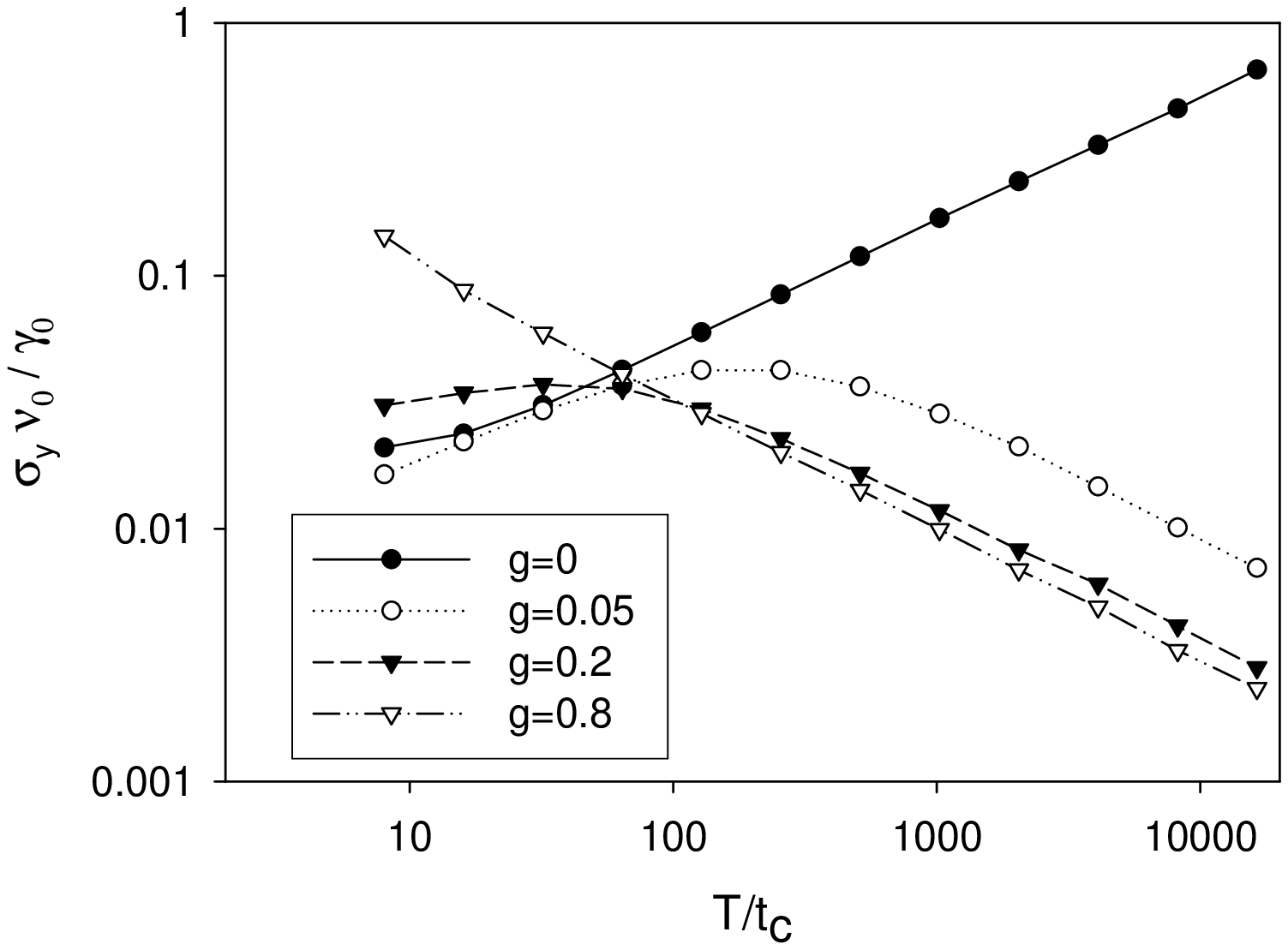}
\end{center}
\caption{Scaled Allan deviation $\sigma_y \nu_0 / \gamma_0$ of the laser frequency calculated from simulated realizations of the servo algorithm for different values of the servo gain $g$. The free-running laser ($g=0$) is assumed to perform a random walk of frequency.}
\end{figure}

The Monte Carlo simulations can also be used to study the instability of the frequency standard that is obtained with this servo algorithm. An excitation spectrum like in equation (13) with $p_{\rm max}=1$ was assumed and laser frequency noise was modeled as a random walk of frequency with maximal step size $\gamma_0/200$ and 10  steps per cycle,
corresponding to conditions where the short-term laser linewidth is smaller than $\gamma_0$, as realized in some experiments \cite{rafac,peik}.
In figure (4) the Allan deviation is plotted for the free-running laser (increasing as $\sqrt{T}$) and for servo parameters $z=4$ and different values of $g$.
For a small gain like $g=0.05$, the short-term instability is mainly determined by the free-running laser. $\sigma_y(T)$ then  reaches a maximum at $T\approx 250 t_c$ or $T\approx 2t_{\rm servo}$ and decreases like $1/\sqrt{T}$ for $T>1000t_c$.
The value $g=0.05$ is too small to completely eliminate the influence of laser noise at long averaging times, as can be seen from the comparison with the curve obtained for $g=0.8$:
Here, the short-term stability is worse than that of the free-running laser, but the long-term stability reaches an optimum that is only limited
by QPN and is independent of $g$, $z$, and the laser frequency noise.
This minimized instability follows the scaling $\sigma_y =0.29(\gamma_0 /\nu_0) \cdot \sqrt{t_c/T}$ for the  excitation spectrum in equation (13) with $p_{\rm max}=1$. We verified numerically that for smaller values of $p_{\rm max}$ the numerical prefactor increases like $\sqrt{(2-p_{\rm max})/p_{\rm max})}$, in agreement with equations (1,7). 
Equation (13) can be used to approximate the shape of the spectrum for Rabi excitation in figure (2) with
parameters $\gamma_0\approx 1/2\tau$ and $p_{\rm max}\approx 0.5$ for $t_c\approx 2\tau$. 
This leads to an Allan deviation $\sigma_y=0.36 /(\nu_0\sqrt{\tau T})$ that can be compared to the scaling obtained in section (3) for $S_P=3.22$:  $\sigma_y=0.51 /(\nu_0\sqrt{\tau T})$. As discussed above, this indicates that the Allan deviations as calculated from $S_P$ and equation (7) are somewhat pessimistic because of the approximation we used for the slope of the spectrum at the FWHM point.

In a practical realization of the algorithm one has to take into account that some cycles may fail to provide  clear information on the effect of the last probe pulse, for example because of momentary heating of the ion
or of an extended sojourn in the metastable level that extends into the following cycle. 
We call these cycles `invalid' for short. In our experimental realization, for symmetry reasons an equal number of interrogations at $f_+$ and $f_-$ is processed in each averaging interval. To achieve this, the sign of the laser detuning is not changed after an invalid cycle, but the cycle is repeated at the same detuning.
If at the end of an averaging interval the numbers of valid cycles at $f_+$ and at $f_-$ are unbalanced the result of the last unpaired valid cycle is discarded.
If invalid cycles occur with a probability $p_{\rm inv}$, the long-term Allan deviation is increased by a factor $1/\sqrt{1-p_{\rm inv}}$ because of the reduced rate of frequency correction information.    

In summary, the results of the Monte Carlo simulations suggest that a small value of cycles $2z$ per averaging interval should be chosen in order to optimize the temporal response of the servo system. Gain values in the range $0.1<g<1$ will provide the best overall stability, depending on the amount and spectral characteristics of the laser frequency noise.

\section{Servo response to laser frequency drift}

A significant servo error may occur if the probe laser frequency is subject to drift. The short-term frequency stability is usually derived from a high-finesse Fabry-Perot cavity made from materials with small coefficients of thermal expansion like ULE or Zerodur. These materials show long-term dimensional changes. Due to this and as a result of residual thermal effects, laser frequency drift rates 
$|df/dt|$ in the range 0.05 Hz/s up to several Hz/s are typically observed. The actual drift rate may change significantly on a time scale of one hour. 
For a conventional first-order integrating servo not operating in discrete time steps, an average drift-induced error $\bar e =t_{\rm servo}\,df/dt$ is expected as the result of a constant linear drift \cite{savant}.
Our numerical simulations yield the same result for the servo system considered here. It is found that the servo error is independent of the temporal order of the interrogations at $f_+$ and $f_-$ within the averaging intervals. Since the minimally achievable servo time constant is 
$t_{\rm servo}\approx 2zt_c$ for stable operation (see above), the drift-induced error can not be reduced to less than $\bar e \approx 2zt_c df/dt$. Thus the output of the frequency standard could differ from the resonance line center by several hertz under unfavorable conditions. 

An efficient reduction of this servo error is obtained with the use of a 
second-order integrating servo algorithm where 
a drift correction $e_{\rm dr}$ is applied to the laser frequency in regular time intervals $t_{\rm dr}$ 
\begin{equation}
f \stackrel{t_{\rm dr}}{\longrightarrow} f+e_{\rm dr}.	
\end{equation}
The drift correction is calculated from the integration 
of the error signal (equation (11)) over a longer
time interval $T_{\rm dr}\gg t_{\rm dr}$ 
\begin{equation}
e_{\rm dr} \stackrel{T_{\rm dr}}{\longrightarrow} e_{\rm dr}+k \sum_{T_{\rm dr}} e.	
\end{equation}
The long averaging time $T_{\rm dr}$ is introduced to reduce the influence of QPN on the drift correction $e_{\rm dr}$. If the two gain coefficients are related by $k\ll g$, the use of the second-order integrating servo algorithm has a negligible influence on the time constant $t_{\rm servo}$. It introduces a second, much longer time constant
that is approximately proportional to $gzt_c/k$.  
In our practical realization (see below) with $t_c\approx 0.1$~s and $|df/dt|<0.3$~Hz/s, we use
$t_{\rm drift}=1$~s, $T_{\rm drift}=10$~s,  and $k$ in the range $3\cdot10^{-4}$--$3\cdot10^{-3}$.
A numerical simulation with the values $g=0.2$, $k=0.002$, $z=4$ and $p_{\rm max}=0.6$ showed two clearly separated servo time constants of approximately $50t_c$ and $6500t_c$.
Ideally, a second-order integrator will completely eliminate the error induced by a linear drift \cite{savant}. 
In the numerical simulations performed for the conditions given above, no statistically significant servo error due to a linear drift was detectable down to the millihertz level.

\section{Comparison with the experiment}    

The servo algorithm described above is experimentally implemented in the $^{171}$Yb$^+$ single-ion optical frequency standard at PTB \cite{peik,tamm1,stenger,schneider}. The reference transition of this standard is 
the electric quadrupole transition $^2S_{1/2}(F=0) \rightarrow {^2D_{3/2}(F=2)}$
at 688 THz (436 nm wavelength) with a natural linewidth of 3.1 Hz. 
A single ytterbium ion is stored in a miniature Paul trap and is laser-cooled to a sub-millikelvin temperature
by exciting the low-frequency wing of the quasi-cyclic hyperfine component 
$^2S_{1/2}(F=1) \rightarrow {^2P_{1/2}}(F=0)$ of the resonance transition at 370~nm. The 
corresponding mean vibrational quantum numbers for the secular oscillations with frequencies of 0.7 MHz and 1.4 MHz are in the range 10--20.
A weak sideband of the cooling radiation provides hyperfine repumping from the $(F=0)$ ground state to the $^2P_{1/2}(F=1)$ level. At the end of each cooling phase, the hyperfine repumping is switched off in order to prepare the ion in the $(F=0)$ ground state. The reference transition is probed by the frequency doubled radiation from a diode laser
emitting at 871 nm. The short-term frequency stability of this laser is derived from a temperature-stabilized and seismically isolated high-finesse cavity made from ULE.
In this work probe pulses with durations ranging from 30--100 ms are used,
leading to linewidths from 10--30 Hz.
From the fact that essentially Fourier-limited resonances with 10 Hz linewidth are obtained, we conclude that the  linewidth of the probe laser is well below 10 Hz \cite{peik}.
The excitation to the metastable $^2D_{3/2}$ level is inferred from the resonance fluorescence level detected 
in the first 5~ms after reapplication of the cooling laser. 
The rate of invalid cycles (see section (5)) can be as high as  30\% for a high excitation probability to the $^2D_{3/2}$ level.
The total cycle dead time needed for laser cooling, hyperfine pumping, and state detection after the probe pulse,  
is presently 60~ms. 

To enable a quantitative study of the instability for long averaging times,
we measure the frequency difference between two similar systems with  single $^{171}$Yb$^+$ ions stored in separate traps \cite{schneider}.
Both traps use the same cooling laser setup and synchronous timing 
for cooling, state preparation, and state detection within each cycle.
The beam from the probe laser is split and  two
separate frequency shift and servo systems are employed to stabilize the probe frequencies to the reference transition line centers of the two ions. The frequency difference between the two systems is recorded once per second.
For times that are larger than $t_{\rm servo}$ the instability of this difference signal is independent of the temporal overlap of the probe pulses in the two systems.  This seems to indicate that the observed long-term instability of the frequency difference is not affected by correlations of the servo responses to short-term laser frequency fluctuations.
Temporal changes of the laser frequency drift rate did not contribute to the observed instability, even when the time constants of the two second-order integrating servos were not matched.
Consequently, the experiment allows us to measure
the combined instability of two independent single-ion frequency standards.

\begin{figure}[t]
\begin{center}
\includegraphics[width=7cm]{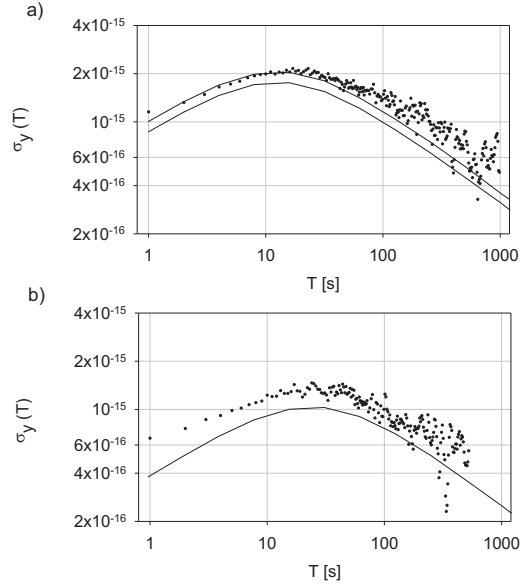}
\end{center}
\caption{Allan deviation of the frequency difference between two Yb$^+$ single-ion optical frequency standards at 688 THz. The experimental data points are compared to Monte Carlo simulations (solid lines) of the QPN limited instability for the experimentally realized servo parameters. The reference transition linewidth is 30 Hz (a) and 10 Hz (b). Two solid curves are plotted in (a) because slightly different servo parameters were used in the two systems (see text).}
\end{figure}

Figure (5a) shows the Allan deviation of the frequency difference for Rabi excitation with $t=0.03$~s and $t_c=0.09$~s. Here the reference transition was resolved with a linewidth of 30 Hz in both traps, so that the influence of the probe laser linewidth was negligible. For averaging times $T>80$~s, the experimental data points approximately follow the scaling $\sigma_y(T)\approx 1.1\cdot 10^{-14}\, /\sqrt{T/{\rm s}}$. Under the assumption of identical behaviour of the two standards, this corresponds to
an instability $\sigma_y(T)\approx 7.8\cdot 10^{-15}\, /\sqrt{T/{\rm s}}$ for the individual system, about a factor of three above the optimal stability that is achievable for idealized conditions (cf. section (3)).
The solid lines show the results of Monte Carlo simulations for the realized servo parameters 
$\delta_m=13$~Hz, $g=0.2$, and $k=0.0005$. 
The shape of the excitation spectrum was obtained from the integration of the Bloch equations for the experimental pulse parameters.
Two curves are presented since the two systems were operated under slightly different conditions:
The curve with the lower Allan deviation is calculated with 
$z=5$, $p_{\rm max}=0.65$ and $p_{\rm inv}=0.30$, while the curve with the higher Allan deviation represents the conditions $z=4$, $p_{\rm max}=0.55$ and $p_{\rm inv}=0.35$. 
According to equation (14) these sets of parameters result in approximately equal servo time constants 
$t_{\rm servo}\approx 7$~s.
Both simulated single-system instabilities were multiplied by $\sqrt{2}$ for comparison with the instability of the frequency difference.

The Monte Carlo simulations assume a monochromatic laser excitation and include QPN as the only source of noise.
It can be seen that the agreement between experiment and simulation is good and that the experimentally achieved instability is essentially consistent with the QPN limit, except for an excess instability of about 20\% at the longest averaging times.

Figure (5b) shows a similar evaluation for a reference transition linewidth of about 10 Hz, obtained with Rabi excitation and $t=0.09$~s, $t_c=0.16$~s. 
According to the optimization shown in figure (1), a linewidth of about 10~Hz will provide the lowest possible instability for the natural linewidth $\Delta\nu_0=3.1$~Hz of the 688 THz reference transition of Yb$^+$.
Both systems were operated under identical conditions: $\delta_m=4$~Hz, $g=0.4$, $z=5$, $p_{\rm max}\approx 0.35$, and
$p_{\rm inv}\approx 0.3$.
The servo time constant estimated from equation (14) is 6~s. 
The experimental result for the instability of the frequency difference is $\sigma_y(100\,{\rm s})=9\cdot10^{-16}$.  
Again, the solid line shows the result of the simulated single-system instability, multiplied by $\sqrt{2}$. Here an excess noise of about 50\% above the QPN limit is observed in the experimental data over the whole range of averaging times.
This stability degradation might be related to probe laser frequency fluctuations and to 
uncorrelated fluctuations of the optical path lengths to the traps.

\section{Conclusion}

We have discussed practical aspects of the problem of the stabilization of a laser frequency using the signal from a single trapped ion. With an experimental implementation in the $^{171}$Yb$^+$ optical frequency standard we have obtained a quantum projection noise limited instability below
$\sigma_y(T)=1\cdot 10^{-14}/\sqrt{T/{\rm s}}$, comparable to the best results obtained so far with optical frequency standards based on ensembles of more than $10^6$ cold Ca atoms \cite{oates} and in a comparison of standards based on
a single Hg$^+$ ion and on a Ca atom cloud \cite{oskay2}.       

Some technical improvements of our standard appear straightforward.
In particular, it would be beneficial for the stability to reduce the rate of invalid cycles. An additional hyperfine pump frequency could drive the $^2S_{1/2}(F=1)\rightarrow {^2P}_{1/2}(F=1)$ component of the resonance transition during the state preparation and a quenching laser may be used to deplete the metastable $^2D_{3/2}(F=2)$ state after an excitation has been detected.
These improvements may also allow us to reduce the dead time to about 20 ms and to approach the optimal Allan deviation of $\sigma_y(T)\approx 2\cdot 10^{-15}/\sqrt{T/{\rm s}}$ that is estimated for this system.

Quantum projection noise imposes a fundamental limitation on the stability of a frequency standard that uses a single ion as a reference. We have discussed how the interrogation parameters have to be chosen in order to obtain the lowest possible long-term instability, taking realistic experimental parameters into account: For a reference transition with natural linewidth $\Delta \nu_0$, a laser linewidth that is smaller than $0.1\Delta\nu_0$, and a cycle dead time $t_d<0.2/(2\pi\Delta\nu_0)$,
an instability $\sigma_y\approx 2 (\Delta \nu_0/\nu_0)\sqrt{1/(2\pi\Delta\nu_0 T)}$ will be achievable.
For an optical frequency $\nu_0 \approx 1\cdot 10^{15}$~Hz, a very narrow linewidth $\Delta\nu_0 \approx 0.01$~Hz would have to be chosen if the instability shall reach the level of $10^{-18}$ 
within a few hours of averaging.  While suitably narrow reference transitions exist in several ions \cite{blythe,schmidt,th}, substantial improvements in laser linewidth narrowing are required to progress into this regime of ultrahigh resolution spectroscopy.

\ack
We thank A. Bauch and R. Wynands for careful reading of the manuscript.
This work was supported by DFG through SFB 407.

\end{document}